\documentclass[aps,prl,showpacs,nofootinbib,twocolumn,float,groupedaddress]{revtex4}
\usepackage{amsmath,bm,epsfig,graphicx}

\begin{document}
\title{Inelastic collisions of optically trapped metastable calcium atoms}
\author{Purbasha Halder, Hannes Winter, and Andreas Hemmerich}
\affiliation{Institut f\"{u}r Laserphysik, Universit\"{a}t Hamburg, Luruper Chaussee 149, 22761 Hamburg, Germany}

\begin{abstract}
We study binary collisions of metastable calcium atoms ($^{40}$Ca) in an optical dipole trap. Collisions between $^{3}$P$_{0}$-atoms and between $^{3}$P$_{0}$ and $^{1}$S$_{0}$-atoms are considered. In the former case, the elastic and inelastic collision parameters are found to be $5.4\times 10^{-11}\,\mathrm{cm}^{3}\mathrm{s}^{-1}$ and $3.6\times 10^{-11}\,\mathrm{cm}^{3}\mathrm{s}^{-1}$, respectively. A fraction of the collisions between $^{3}$P$_{0}$-atoms is found to produce cold trapped atoms in the singlet $^{1}$S$_{0}$ state, suggesting that the internal energy for these collisions is dissipated by radiation. For collisions between $^{3}$P$_{0}$ and $^{1}$S$_{0}$-atoms we find a two-body loss parameter of $8.5\times 10^{-11}\,\mathrm{cm}^{3}\mathrm{s}^{-1}$. Our observations show that metastable calcium samples in the $^{3}$P$_{0}$-state are not stable at high densities, as for example required in quantum computing or many-body quantum simulation schemes.
\end{abstract}

\date{\today}

\maketitle
The collision properties of alkaline-earth metal (AEM) and rare-earth metal (REM) atoms in their metastable states are of fundamental interest in different areas of atomic physics. For example, precise knowledge of collisional shifts of their narrow band intercombination transitions is required in atomic clock applications~\cite{Lis2009}. The recently proposed implementation of quantum computing and quantum simulation schemes using metastable triplet states~\cite{Dal2008, Fos2010, Gor2010} rely on suitable control of collisional relaxation, as does the prospect of achieving quantum degeneracy in these states.

In previous experiments two-body scattering rates in various trap geometries for various metastable states of AEM- and REM-atoms, including magnetically trapped calcium atoms in the $^{3}$P$_{2}$-state ~\cite{Han2006}, optically trapped ytterbium atoms in the $^{3}$P$_{2}$-state~\cite{Yam2008, Uet2012} and strontium atoms in the $^{3}$P$_{0}$- and $^{3}$P$_{2}$-states~\cite{Tra2009} have been determined. On the theoretical side, there have been efforts to understand and predict collisional properties of metastable states from first principles ~\cite{Der2003, Kok2003, San2003}. Various models have been used to analyze trap decay data in order to deduce scattering rates~\cite{Spo2005, Lis2009, Yan2011}. Collision studies of magnetically trapped $^{3}$P$_{2}$ calcium atoms have shown that elastic scattering rates are large (around $3\times10^{-10}$\,cm$^{3}$s$^{-1}$) well above the $s$-wave unitarity limit and that inelastic scattering rates are of similar magnitude \cite{Han2006}. This supported theoretical predictions that higher order collisional channels strongly modify the behaviour of metastable Ca and Sr atoms during a scattering event \cite{Der2003, Kok2003, San2003}. The large inelastic collisions rates have been attributed to magnetic quantum number changing and fine structure changing collisions. Such collisions may be avoided by using an optical dipole trap (ODT), confining the atoms in the non-magnetic $^{3}$P$_{0}$-state, which represents the ground state of the triplet manifold. Nevertheless, experiments with strontium in an ODT \cite{Yan2011} have shown that even in the $^{3}$P$_{0}$-state significant collisional relaxation is observed. Significant relaxation rates were also observed for collisions between ytterbium $^{3}$P$_{2}$-and $^{1}$S$_{0}$-atoms \cite{Yam2008, Uet2012}, which is equally relevant for atomic clock applications and quantum simulation schemes.

To date there has been no report of similar work for optically trapped calcium atoms. The reason may be that laser cooling and trap loading schemes used successfully for Sr and Yb cannot be easily applied to the case of calcium, since the linewidth of the $^{1}$S$_{0} \rightarrow ^{3}$P$_{1}$ intercombination transition is too narrow to allow efficient cooling unless sophisticated line quenching and laser bandwidth shaping techniques are employed \cite{Bin2001, Deg2005}. In this work we apply an alternative loading technique, previously used to produce a calcium Bose-Einstein condensate (BEC) in the singlet ground state \cite{Hal2012,Yan2007}, to prepare samples of metastable calcium atoms (bosonic $^{40}$Ca) in an ODT in the $^{3}$P$_{0}$ state with sufficient density to enable collision studies. This has allowed us to study binary collisions between $^{3}$P$_{0}$-atoms and between $^{3}$P$_{0}$ and $^{1}$S$_{0}$-atoms. In the former case, the elastic and inelastic collision parameters are found to be $5.4\times 10^{-11}\,\mathrm{cm}^{3}\mathrm{s}^{-1}$ and $3.6\times 10^{-11}\,\mathrm{cm}^{3}\mathrm{s}^{-1}$, respectively. The comparatively large inelastic collision parameter excludes the possibility of efficient evaporative cooling in the $^{3}$P$_{0}$-state. A fraction of the collisions between $^{3}$P$_{0}$-atoms is found to produce cold trapped atoms in the $^{1}$S$_{0}$ state. For these collisions the internal energy can only be dissipated by emission of red photons. For collisions between $^{3}$P$_{0}$ and $^{1}$S$_{0}$-atoms we find a two-body loss parameter of $8.5\times 10^{-11}\,\mathrm{cm}^{3}\mathrm{s}^{-1}$. This large value indicates that even the lifetime of dilute $^{3}$P$_{0}$-samples is significantly limited by the presence of a dense ($> 10^{11} \mathrm{cm}^{-3}$) $^{1}$S$_{0}$-background. We conclude that (without additional measures) metastable calcium samples in the $^{3}$P$_{0}$-state are not suitable for implementing scenarios, which require high densities, as for example quantum computing or many-body quantum simulation schemes \cite{Dal2008, Fos2010, Gor2010}.

\begin{figure}
\centering
\includegraphics[width=86mm]{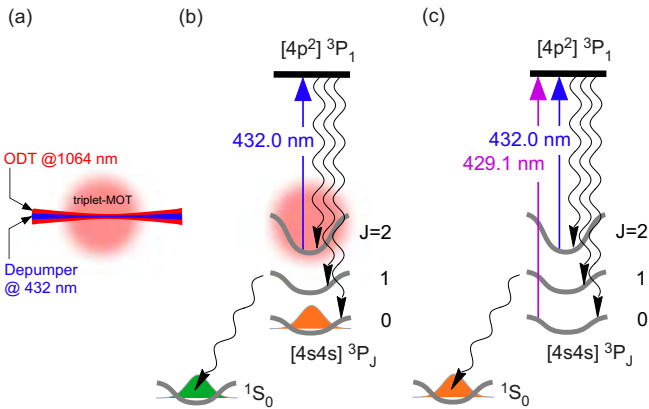}
\caption{(Color online) (a) Sketch of the optical dipole trap (ODT) and 432.0\,nm depumping laser beams (both not to scale) in the triplet-MOT; gravity acts perpendicular to the ODT. (b) Depumping scheme for loading ODT with $^{3}$P$_{0}$-atoms. (c) Scheme for depumping $^{3}$P$_{0}$-atoms to the ground state for imaging. Two lasers at 432.0\,nm and 429.1\,nm are used for this purpose. In (b) and (c), straight lines denote laser transitions and wavy lines denote spontaneous decay.}
\vspace{6.mm}
\label{Fig.1}
\end{figure}

\begin{figure}[t!]
\centering
\includegraphics[width=86mm]{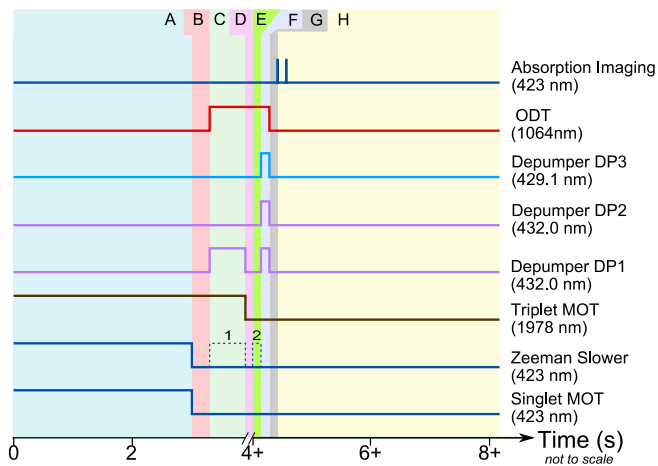}
\caption{Experimental sequence. A. (3\,s)\,Loading of triplet MOT; B. (300\,ms)\,Cooling in the triplet-MOT; C. (600\,ms)\,Loading of dipole trap with $^{3}\mathrm{P}_{0}$ atoms; D. (\textit{variable})\,Holding time in the trap; E. (\textit{optional}, 2.5\,ms)\,Removal of remaining ground state atoms; F. (1\,ms)\,Depumping atoms to the ground state; G. (\textit{variable})\,time-of-flight; H. ($\sim 4$\,s)\,Imaging and image processing. To clean the dipole trap of ground state atoms, the Zeeman slower beam may be kept on either in time window 1 or 2, depending on whether we wish to observe collisions of $^{3}\mathrm{P}_{0}$ atoms without or in presence of ground state atoms, respectively.}
\label{Fig.2}
\end{figure}

For all experiments reported here, the ODT beam is derived from a 20\,W fiber laser (IPG photonics, YLR series) at 1064\,nm. With a beam waist of 31\,$\mu$m, a power of 13.5\,W inside the vacuum chamber and $\sigma^{+}$ polarized light, the resulting maximum trap depth is $270\,\mu$K for the $^{3}\mathrm{P}_{0}$ state, $644\,\mu$K for the $m=-2$ sublevel of the $^{3}\mathrm{P}_{2}$ state and 370\,$\mu$K for the $^{1}\mathrm{S}_{0}$ ground state. Our scheme for loading the ODT with metastable atoms is based on the effect of spatially selective depumping previously used in our group for preparing dense, cold samples of $^{1}\mathrm{S}_{0}$ ground state atoms \cite{Yan2007}, and subsequently a $^{1}\mathrm{S}_{0}$ BEC \cite{Hal2012}. In this scheme, a weak and tightly focussed depumping beam (DP1) is adjusted to lie well within the volume of the ODT beam (see Fig.\,~\ref{Fig.1}(a)). Both beams propagate through the centre of a magneto-optical trap (termed triplet-MOT) operating on the infrared $^{3}P_{2}\, \rightarrow \, ^{3}D_{3}$ transition at 1978\,nm, thus preparing cold atoms in the $^{3}P_{2}$-state \cite{Gru2002, Han2003}. The triplet-MOT receives pre-cooled atoms from a spatially superimposed magneto-optic trap (singlet-MOT) collecting ($^{1}\mathrm{S}_{0}$) singlet atoms previously decelerated in a Zeeman slower. For loading the ODT with $^{3}P_{0}$-atoms, DP1 is chosen to operate at 432.0\,nm resonant with the $[4s4p] ^{3}P_{2}\, \rightarrow [4p^{2}]^{3}P_{1}$ transition (Fig.\,~\ref{Fig.1}(b)). Atoms from the excited state of the depumping transition decay to all the three levels in the $[4s4p] ^{3}P_{J}$ triplet-manifold with branching ratios 33\%, 26\% and 41\% corresponding to $J=0,1$ or 2, respectively. The atoms in the $^{3}P_{2}$ state are recycled, while the $^{3}P_{1}$ state eventually decays to the ground state ($^{1}S_{0}$) in about 380\,$\mu$s \cite{Deg2005}. DP1 has a power of a few 10\,nW and its waist is 17\,$\mu$m.

The experimental sequence followed in our studies of metastable $^{3}$P$_{0}$ atoms is illustrated in Fig.\,~\ref{Fig.2}. Stages A. and B. are identical to step\,1 of the protocol described in \cite{Hal2012}, where atoms are prepared in the $^{3}P_{2}$ state in the triplet-MOT. In the 600\,ms dipole trap loading stage C., the ODT and DP1 beams are switched on. In the region of the dipole trap, triplet-MOT atoms are thus continuously transferred to the $^{3}P_{0}$ and $^{1}S_{0}$ states, both of which are trapped in the optical potential. For collision studies in the absence of ground state atoms, the Zeeman slower beam operating close to the $^{1}S_{0}\, \rightarrow\, ^{1}P_{1}$ transition illuminates the MOT region during this time and \textit{blows away} the trapped ground state atoms. If this \textit{cleaning} beam is inactive during stage C., both the $^{3}P_{0}$ and $^{1}S_{0}$ levels are simultaneously populated. We observe that about twice as many ground state atoms (65\%) as metastable ones (35\%) are loaded, although the branching ratio predicts a $44:56$ distribution. We attribute this discrepancy to the lower trap depth and the large two-body collision loss rate for the metastable atoms.

The atoms collected in stage C. are held in the trap for a variable duration during stage D. before an absorption image is taken. For determining collision rates in the presence of the ground state atoms, these were not \textit{blown away} in step C. This amounts to observing $^{3}P_{0}$ collisions influenced by the presence of a dense cloud of ground state atoms. In this case, an additional \textit{cleaning} pulse applied during stage E. of 2.5\,ms duration gets rid of the ground state atoms before imaging. The atoms in state  $^{3}P_{0}$ are prepared for absorption imaging by depumping them to the ground state in stage F. This is done by simultaneously applying two laser beams well covering the entire ODT volume. These beams, which we refer to as DP2 and DP3, operate at 432.0\,nm and 429.1\,nm (Fig.\,~\ref{Fig.1}(c)), respectively. DP2 is necessary for keeping the $^{3}P_{2}$ state depopulated. The beams have high intensity ($\sim$ saturation intensity of the transition in either case) such that in less than 1\,ms all atoms are transferred to the ground state. The ODT laser is kept running during this time such that the depumped atoms are also kept confined by the dipole trap potential. This is necessary, since otherwise the relatively hot ensemble of free atoms will expand rapidly, and the optical density will be too low for the atoms to be imaged. During stage G. the dipole trap is switched off and the atoms that have been released from the trap undergo ballistic expansion. The ensemble temperature can be estimated from a series of such time-of-flight measurements. In the final stage H. an absorbtion image of the atoms is produced and the data are processed.

\begin{figure}[t!]
\centering
\includegraphics[width=86mm]{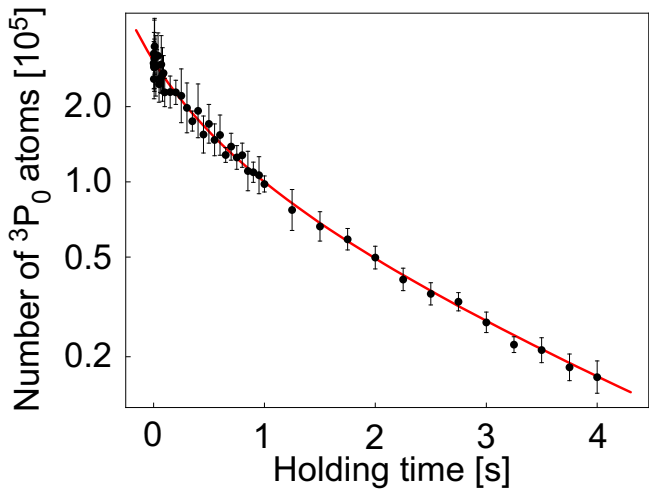}
\caption{Trap decay for a pure sample of optically trapped $^{3}\mathrm{P}_{0}$ atoms. The time $t=0$ corresponds to the end of the loading period (stage C in Fig\,~\ref{Fig.2}). The abscissa corresponds to the variable times in stage D of Fig.\,~\ref{Fig.2}. The data points (black circles) plotted on a logarithmic scale clearly show a non-exponential decay. The solid line shows a fit of the data using Eq.\,~\ref{decay_e}.}
\label{Fig.3}
\end{figure}

To model trap dynamics including the states $|g\rangle =  \,^{1}\mathrm{S}_{0}$ and $|e\rangle = \,^{3}\mathrm{P}_{0}$ we apply a rate equation analysis based upon Refs. \cite{Gru2001}, \cite{Tra2009}, \cite{Lis2009} for the particle numbers $N_{\mathrm{e}}(t)$ and $N_{\mathrm{g}}(t)$ in the states 
$|e\rangle$ and $|g\rangle$
\begin{equation}
\dot{N}_{\mathrm{e}}= R_{\mathrm{e}} - \Gamma_{\mathrm{e}} N_{\mathrm{e}} - \beta_{\mathrm{e}}\,  \frac{V_{\mathrm{e},2}}{V_{\mathrm{e},1}^{2}}\, N_{\mathrm{e}}^{2} - \beta_{\mathrm{eg}}\, \frac{1}{V_{\mathrm{eg}}}\, N_{\mathrm{e}}\,N_{\mathrm{g}}
\label{e_evolution}
\end{equation}
\begin{equation}
\dot{N}_{\mathrm{g}}= R_{\mathrm{g}} -\Gamma_{\mathrm{g}} N_{\mathrm{g}} -L_{\mathrm{g}}\, \frac{V_{\mathrm{g},3}}{V_{\mathrm{g},1}^{3}} \,N_{\mathrm{g}}^{3} - \beta_{\mathrm{eg}} \frac{1}{V_{\mathrm{eg}}} N_{\mathrm{e}}\,N_{\mathrm{g}}
\label{g_evolution}
\end{equation}
with $\Gamma_{\mathrm{a}}$, $a \in \{ \mathrm{e},\mathrm{g} \}$ denoting the linear loss rates due to collisions with background atoms, $R_{\mathrm{a}}$ with $a \in \{ \mathrm{e},\mathrm{g} \}$ denoting the loading rates via optical depumping from the triplet-MOT, and $\beta_{\mathrm{eg}}$ denoting the loss parameter due to collisions between $|e\rangle$- and $|g\rangle$-atoms. Two-body collisional loss does only arise for $|e\rangle$-atoms parametrized by $\beta_{\mathrm{e}}$. The relatively large ground-state scattering length of calcium ($\sim 440 a_0$, \cite{Kra2009}) comes with a large cross section for three-body recombination, which is accounted for by the loss parameter $L_{\mathrm{g}}$. The effective volumes $V_{\mathrm{a},q}$ are given by
\begin{equation}
V_{\mathrm{a},q}= \frac{1}{n_{\mathrm{a},0}^{q}} \int n_{\mathrm{a}}(\mathbf{r})^{q} d^{3}r
\label{effective_volume}
\end{equation}
for integer $q$ and $n_{\mathrm{a}}(\mathbf{r})$, $n_{\mathrm{a},0}$ with ($a \in \{ \mathrm{e},\mathrm{g} \}$) denoting the density distribution and the peak density for $|a\rangle$-atoms. Finally, the effective volume for interspecies collisions is
\begin{equation}
\frac{1}{V_{\mathrm{eg}}}  = \frac{1}{n_{\mathrm{e},0} \, V_{\mathrm{e},1} \, n_{\mathrm{g},0} \, V_{\mathrm{g,1}}}  \, \int n_{e}(\mathbf{r})\,n_{g}(\mathbf{r}) d^{3}r    ,
\label{effective_volume_eg}
\end{equation}
Making use of the known trap geometry with the trap edges given by the $1/e^{2}$ trap depth and assuming Boltzman distributions for $n_{e}(\mathbf{r})$ and $n_{g}(\mathbf{r})$, Eqs.\,~\ref{effective_volume} and \ref{effective_volume_eg} let us calculate the effective volumes as functions of the temperature. If temperature variations due to evaporative and sympathetic cooling can be neglected the effective volumes are independent of time. 

\begin{figure}
\centering
\includegraphics[width=86mm]{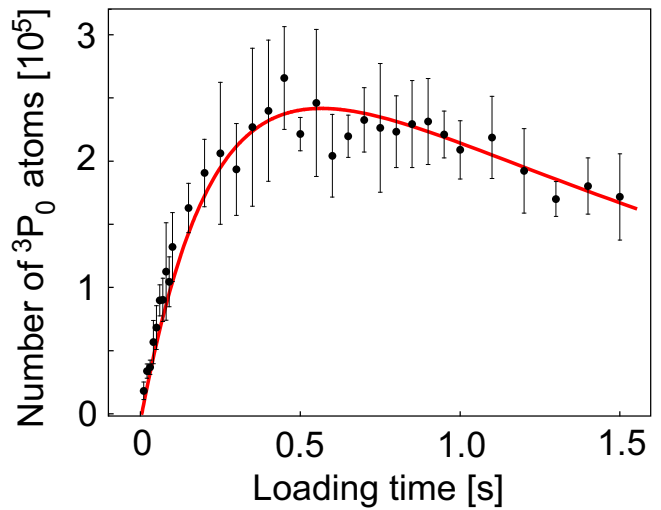}
\caption{(Color online) A typical ODT loading curve. The data points represent the number of $^{3}\mathrm{P}_{0}$-atoms captured in the ODT from the triplet-MOT. The abscissa shows the time for which the depumping laser beam (DP1) is kept on (duration of stage C. in Fig.\,~\ref{Fig.2}).}
\label{Fig.4}
\end{figure}

Fig.\,~\ref{Fig.3} analyses the decay of a pure sample of $^{3}\mathrm{P}_{0}$-atoms after loading has been terminated ($R_{\mathrm{e}}=0$). To model the data, we solve Eq.\,~\ref{e_evolution} with $N_{\mathrm{g}}=0$, $R_{\mathrm{e}}=0$ and with a time-independent linear loss rate $\Gamma_{\mathrm{e}} = \Gamma_{\mathrm{e,0}}$ accounting for collisions of trapped $^{3}P_{0}$-atoms with hot background atoms. This leads to
\begin{eqnarray}
\frac{N_{\mathrm{e}}(t)}{N_{\mathrm{e},0}} = \frac{e^{-\Gamma_{\mathrm{e,0}} t}}{1 + \frac{\beta'_{\mathrm{e}}}{\Gamma_{\mathrm{e,0}}} N_{\mathrm{e},0} \, \left(1-e^{-\Gamma_{\mathrm{e,0}} t}\right)}\,,\,\, \beta'_{\mathrm{e}} \equiv 
\beta_{\mathrm{e}} \,\frac{V_{\mathrm{e},2}}{V_{\mathrm{e},1}^{2}}
\label{decay_e}
\end{eqnarray}
with $N_{\mathrm{e},0}$ denoting the initial number of loaded atoms. The trap decay data in Fig.\,~\ref{Fig.3} are well approximated by the solid red line resulting from Eq.\,~\ref{decay_e} with fitted parameters $\Gamma_{\mathrm{e,0}} = 0.42\,$s$^{-1}$ and $\beta'_{\mathrm{e}} = 4.0\times 10^{-6}\,\mathrm{s}^{-1}$, which yields $\beta_{\mathrm{e}} = 4.3\times 10^{-11}\,\mathrm{cm}^{3}\mathrm{s}^{-1}$. Using the model of Ref. \cite{Car2004} for a linear trap with a large truncation factor $\eta > 4$, the elastic and inelastic collision rates are calculated to be $5.4\times 10^{-11}\,\mathrm{cm}^{3}\mathrm{s}^{-1}$ and $3.6\times 10^{-11}\,\mathrm{cm}^{3}\mathrm{s}^{-1}$ respectively.

The loading process in case of a pure $^{3}\mathrm{P}_{0}$-sample is shown in Fig.\,~\ref{Fig.4}. The loading time and the power of DP1 were optimized in order to maximize the number of atoms in the ODT. The observed decrease in the number of atoms after several hundred ms is attributed to the depletion of the triplet-MOT, which has a background gas limited lifetime. We typically obtain $3.0\times 10^{5}$ $^{3}P_{0}$ atoms after 600\,ms of loading, at a peak density of $1.4\times10^{11}$\,cm$^{-3}$ and a temperature of $64\,\mu$K. The phase space density (PSD) of $5.5 \times 10^{-6}$ is a factor 50 higher than that in the MOT, but still about two orders of magnitude lower than the PSD of ground state atoms at the end of the loading stage in the dipole trap. The data are modeled by Eq.\,~\ref{e_evolution} setting $N_{\mathrm{g}}=0$, $R_{\mathrm{e}} = R_{\mathrm{e},0} \,e^{- \Gamma_{\mathrm{MOT}}}$ and $\Gamma_{\mathrm{e}} = \Gamma_{\mathrm{e,0}} + \Gamma_{\mathrm{e,1}} \,e^{- \Gamma_{\mathrm{MOT}}}$. Here, $R_{\mathrm{e},0}$ is the initial loading rate and $\Gamma_{\mathrm{MOT}}$ is the MOT decay rate. Collisions of trapped $^{3}P_{0}$-atoms with hot background atoms and MOT atoms are accounted for by the rates $\Gamma_{\mathrm{e,0}}$ and $\Gamma_{\mathrm{e,1}}$, respectively. The initial loading rate $R_{\mathrm{e},0}$ is found to be $1.6 \times 10^6$\,s$^{-1}$ by fitting the first 8 data points by a straight line. This is about six times lower than the maximum capture rates achieved for ground state atoms in a dipole trap operating with 532\,nm radiation \cite{Yan2007}. With the fixed parameter values $\Gamma_{\mathrm{e,0}}=0.42$\,s$^{-1}$ and $\beta'_{\mathrm{e}} = 4.0\times 10^{-6}\,\mathrm{s}^{-1}$, found in in Fig.\,~\ref{Fig.3}, and the condition of an initially empty trap ($N_{\mathrm{e}}(0) = 0$), Eq.\,~\ref{e_evolution} is numerically solved and the parameters $\Gamma_{\mathrm{e,1}}$ and $\Gamma_{\mathrm{MOT}}$ are fitted to yield the red solid line in Fig.\,~\ref{Fig.4} for $\Gamma_{\mathrm{e,1}}=2.12$\,s$^{-1}$ and $\Gamma_{\mathrm{MOT}} = 1.6$\,s$^{-1}$.

\begin{figure}[h!]
\centering
\includegraphics[width=86mm]{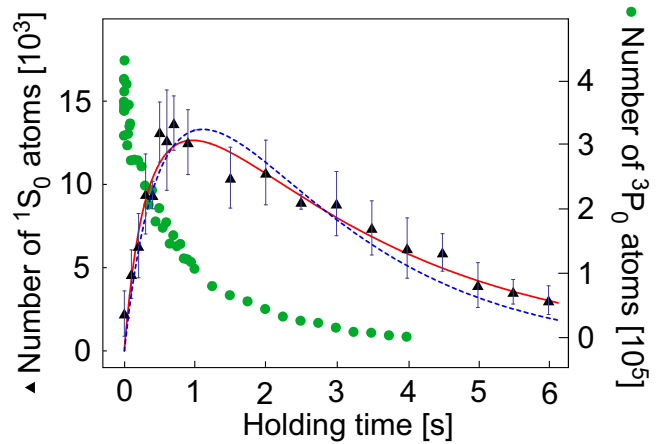}
\caption{Production of trapped ground state atoms (black triangles) resulting from decay of $^{3}\mathrm{P}_{0}$-atoms (green disks). The dashed blue line and the solid red line correspond to fit models based upon Eq.\,~\ref{e_model} and Eq.\,~\ref{g_model}.}
\label{Fig.5}
\end{figure}

In a radiation-free binary inelastic collision between initially slow $^{3}\mathrm{P}_{0}$-atoms the internal energy of at least one of the atoms is equally transferred to the external degrees of freedom of both partners. For example, for the collision channel $^{3}$P$_{0} + ^{3}$P$_{0} \rightarrow ^{1}$S$_{0}  + ^{3}$P$_{0}$ this amounts to 1.89~eV, such that each of the atoms is accelerated to a velocity of about $2100\,$m/s and hence ejected from the trap (which exhibits a small escape velocity on the order of $0.3\,$m/s). Surprisingly, we find that a small but notable fraction (about 4\%) of the atoms leaving the $^{3}\mathrm{P}_{0}$-trap population remains trapped, however, in the singlet ground state. In Fig.\,~\ref{Fig.5} we see that, as the $^{3}\mathrm{P}_{0}$-sample decays (green disks), the number of ground state atoms grows (black triangles). We attribute this to the fact that the metastable $^{3}\mathrm{P}_{0}$-state is quenched during collisions, such that the internal energy is removed via emission of a photon. The radiation driven portion of the decay rate of $^{3}\mathrm{P}_{0}$-atoms is a direct measure of the collisional broadening of the $^{3}\mathrm{P}_{0} \rightarrow\, ^{1}\mathrm{S}_{0}$ transition linewidth, one of the notorious limitations in atomic clock applications \cite{Lis2009}.

\begin{figure}[t!]
\centering
\includegraphics[width=86mm]{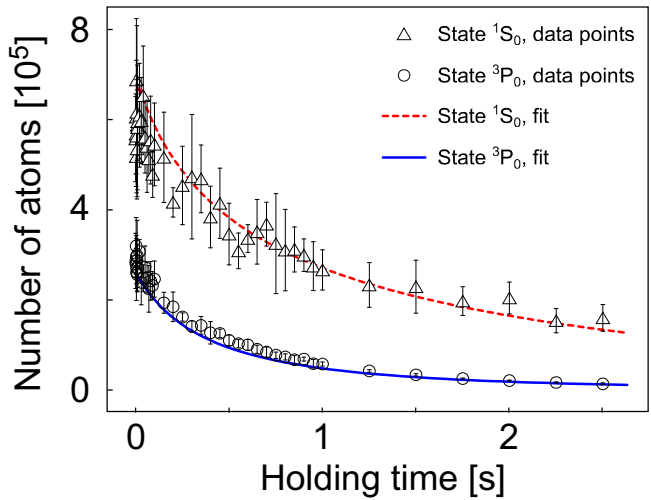}
\caption{Decay dynamics of the states $^{3}\mathrm{P}_{0}$ and $^{1}\mathrm{S}_{0}$ in a sample containing a mixture of both. The line graphs result from Eq.\,~\ref{e_evolution} and Eq.\,~\ref{g_evolution}.}
\label{Fig.6}
\end{figure}

\begin{figure}[t!]
\centering
\includegraphics[width=86mm]{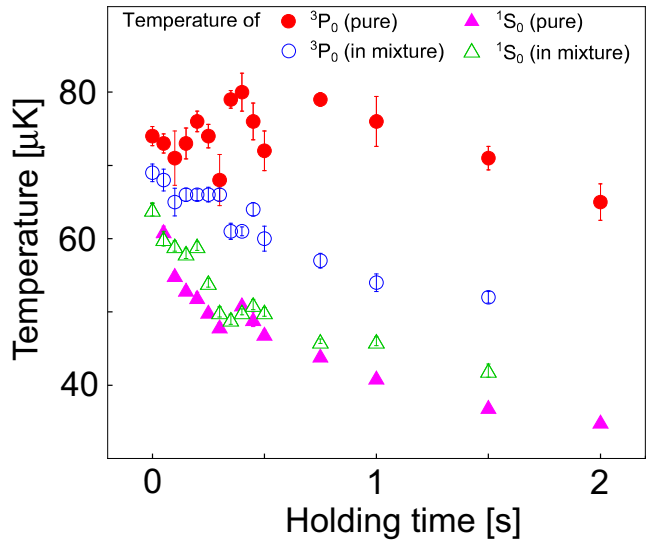}
\caption{Temperature evolution of the relevant atomic states $^{1}\mathrm{S}_{0}$ and $^{3}\mathrm{P}_{0}$ for pure samples or in a mixture.}
\label{Fig.7}
\end{figure}

We now consider two alternative models in order to better understand the production of $^{1}\mathrm{S}_{0}$-atoms in Fig.\,~\ref{Fig.5}. In both cases we assume that the internal energy is removed by radiation, however, in \textsf{model\,1} (\textsf{model\,2}) we assume that this process does not depend (does depend) on the density of $^{3}\mathrm{P}_{0}$-atoms. We hence employ the rate equations
\begin{equation}
\dot{N}_{\mathrm{e}} = - (\Gamma_{\mathrm{e}} + \Gamma_{\mathrm{rad}}) N_{\mathrm{e}} - (\beta'_{\mathrm{e}} + \beta'_{\mathrm{rad}})\, N_{\mathrm{e}}^{2} 
\label{e_model}
\end{equation}
\begin{equation}
\dot{N}_{\mathrm{g}}= -\Gamma_{\mathrm{g}} N_{\mathrm{g}} + \Gamma_{\mathrm{rad}} N_{\mathrm{e}} + \beta'_{\mathrm{rad}}\, N_{\mathrm{e}}^{2} \, .
\label{g_model}
\end{equation}
The parameters $\beta'_{\mathrm{rad}}$ and $\Gamma_{\mathrm{rad}}$ describe the density dependent or density independent radiative conversion of $|e\rangle$- into $|g\rangle$-atoms. We assume $\beta'_{\mathrm{rad}}=0$ in \textsf{model\,1} and $\Gamma_{\mathrm{rad}}=0$ in \textsf{model\,2}. Note that in either case we have neglected the possibility of three-body recombination for singlet atoms, assuming that the population and hence the density in this state remains small. Solving Eq.\,~\ref{e_model} yields
\begin{equation}
\frac{N_{\mathrm{e}}(t)}{N_{\mathrm{e},0}} =\frac{e^{-(\Gamma_{\mathrm{e}} +\Gamma_{\mathrm{rad}})t}}{1+(\frac{\beta'_{\mathrm{e}}+\beta'_{\mathrm{rad}}}{\Gamma_{\mathrm{e}} +\Gamma_{\mathrm{rad}}} ) N_{\mathrm{e},0} (1-e^{-(\Gamma_{\mathrm{e}} +\Gamma_{\mathrm{rad}})})},
\label{e_model_solution}
\end{equation}
Upon inserting Eq.\,~\ref{e_model_solution} into Eq.\,~\ref{g_model} we may numerically solve Eq.\,~\ref{g_model} with the initial condition $N_{\mathrm{g}}(t=0) = 0$ in order to model the data in Fig.\,~\ref{Fig.5}. The values of $\beta'_{\mathrm{e}}$ and $\Gamma_{\mathrm{eg}}$ are taken from the trap decay measurement of Fig.\,~\ref{Fig.3}. If \textsf{model\,1} is used, the best match to the data, shown as the dashed blue line in Fig.\,~\ref{Fig.5}, is obtained with $\Gamma_{\mathrm{rad}} = 0.102 \,\mathrm{s}^{-1}$, $\Gamma_{\mathrm{g}}=0.715\, \mathrm{s}^{-1}$. Alternatively, \textsf{model\,2} leads to the solid red line in Fig.\,~\ref{Fig.5} with optimized fit parameter values $\beta'_{\mathrm{rad}}= 3.6 \times 10^{-7} \,\mathrm{s}^{-1}$ and $\Gamma_{\mathrm{g}}=0.34 \, \mathrm{s}^{-1}$. We conclude that \textsf{model\,2} more accurately matches with the observations, thus confirming, that we observe radiative decay of $^{3}\mathrm{P}_{0}$-atoms via a density dependent collisional quench. This is consistent with the fact that excessive external electric or magnetic fields would be required to yield density-independent line quenching of the observed order of magnitude \cite{Bar2006}.

Finally, we consider a mixture of trapped triplet and singlet atoms. In Fig.\,~\ref{Fig.6} the decay of $^{3}\mathrm{P}_{0}$ and $^{1}\mathrm{S}_{0}$ atoms are shown by the open disks and triangles, respectively. The line graphs result from Eq.\,~\ref{e_evolution} and Eq.\,~\ref{g_evolution} with the fit parameter $\beta_{\mathrm{eg}} = 8.5\times 10^{-11}\,\mathrm{cm}^{3}\mathrm{s}^{-1}$. The value of $L_{\mathrm{g}} = 7.96 \times 10^{-22}$cm$^6$s$^{-1}$ has been determined by trap decay observations for pure $^{1}\mathrm{S}_{0}$ samples. Effects of evaporation in the singlet state, sympathetic cooling and sympathetic evaporation have been neglected in this analysis. This appears justified from observations in Fig.\,~\ref{Fig.7} showing that the temperature decrease for pure $^{1}\mathrm{S}_{0}$-atoms, pure $^{3}\mathrm{P}_{0}$-atoms, as well as a mixture of the two is only moderate.

In conclusion we have measured collision parameters of bosonic calcium atoms at ultralow temperatures, considering binary collisions of $^{3}\mathrm{P}_{0}$ atoms and of $^{3}\mathrm{P}_{0}$ atoms with $^{1}\mathrm{S}_{0}$ atoms. These collision processes are shown to possess a significant inelastic contribution, yielding excessive trap loss at densities above $10^{11}$cm$^{-3}$, similar to previous observations for strontium. This prevents long coherence times even at moderate densities as, for example, required in quantum computing or many-body quantum simulation schemes \cite{Dal2008,Fos2010}.

\begin{acknowledgments}
This work was partially supported by DFG (He2334/9-1). We acknowledge useful discussions with Robin Santra.
\end{acknowledgments}

\end{document}